\documentclass[fleqn,10pt]{SelfArx} % Document font size and equations flushed left

\definecolor{color1}{RGB}{0,0,90} % Color of the article title and sections
\definecolor{color2}{RGB}{0,20,20} % Color of the boxes behind the abstract and headings

\usepackage{hyperref} % Required for hyperlinks
\hypersetup{hidelinks,colorlinks,breaklinks=true,urlcolor=color2,citecolor=color1,linkcolor=color1,bookmarksopen=false,pdftitle={Title},pdfauthor={Author},urlcolor=blue}

\usepackage{graphicx}
\usepackage{natbib}
\usepackage{amsmath}
\usepackage{multirow}
\usepackage{subfigure}

\newcommand{\avenotxt}[1]{<\hspace{-0.1cm}B^{#1}\hspace{-0.1cm}>}
\newcommand{\avetxt}[2]{<\hspace{-0.1cm}B^{#1}_{\mathrm{#2}}\hspace{-0.1cm}>}
\newcommand{\avenoexp}[0]{<\hspace{-0.1cm}B\hspace{-0.1cm}>}

\newcommand{\inte}{\textrm{\scriptsize{in}}}
\newcommand{\exte}{\textrm{\scriptsize{ex}}}

\JournalInfo{Published in Journal of Applied Physics, Vol. 104, 013910, 2008} % Journal information
\Archive{\href{http://dx.doi.org/10.1063/1.2952537}{DOI: 10.1063/1.2952537}} % Additional notes (e.g. copyright, DOI, review/research article)

\PaperTitle{Optimization and improvement of Halbach cylinder design} % Article title

\Authors{R. Bj\o{}rk, C. R. H. Bahl, A. Smith and N. Pryds} % Authors
\affiliation{\textit{Department of Energy Conversion and Storage, Technical University of Denmark - DTU, Frederiksborgvej 399, DK-4000 Roskilde, Denmark}} % Author affiliation
\affiliation{*\textbf{Corresponding author}: rabj@dtu.dk} % Corresponding author

 % Keywords - if you don't want any simply remove all the text between the curly brackets
\newcommand{\keywordname}{Keywords} % Defines the keywords heading name

\Abstract{In this paper we describe the results of a parameter survey of a 16 segmented
    Halbach cylinder in three dimensions in which the parameters internal radius,
    $r_{\inte}$, external radius, $r_{\exte}$, and length, $L$, have been varied. Optimal
    values of $r_{\exte}$ and $L$ were found for a Halbach cylinder with the least possible
    volume of magnets with a given mean flux density in the cylinder bore.
    The volume of the cylinder bore could also be significantly increase by only
    slightly increasing the volume of the magnets, for a fixed mean flux density.
    Placing additional blocks of magnets on the end faces of the Halbach
    cylinder also improved the mean flux density in the cylinder bore, especially so for
    short Halbach cylinders with large $r_{\exte}$. Moreover magnetic cooling as an
    application for Halbach cylinders was considered. A magnetic cooling quality parameter,
    $\Lambda_{\mathrm{cool}}$, was introduced and results showed that this parameter was
    optimal for long Halbach cylinders with small $r_{\exte}$. Using the previously
    mentioned additional blocks of magnets can improve the parameter by as
    much as 15\% as well as improve the homogeneity of the field in the cylinder bore.}

\begin{document}

\flushbottom % Makes all text pages the same height

\maketitle % Print the title and abstract box

%\tableofcontents % Print the contents section

\thispagestyle{empty} % Removes page numbering from the first page

\section{Introduction}\label{Sec.Introduction}
Configurations of permanent magnets that produce a strong homogeneous field in a
confined region of space and a very weak field elsewhere are useful in many applications
such as particle accelerators \cite{Sullivan_1998}, nuclear magnetic resonance (NMR)
apparatus \cite{Appelt_2006} or magnetic cooling applications \cite{Coey_2002}.
%--- Corrected NMR acronym

\begin{figure}[!t]
  \centering
   \includegraphics[width=0.6\columnwidth]{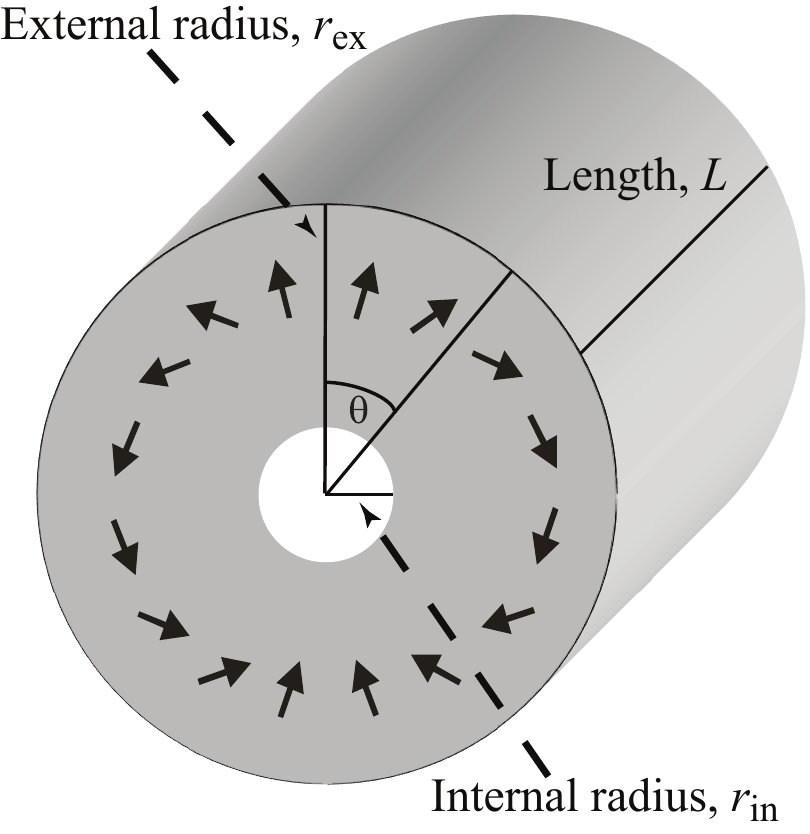}
      \caption{A sketch of a Halbach cylinder showing the internal radius, $r_{\inte}$, external radius, $r_{\exte}$, and length, $L$.
                   Also shown as arrows is the direction of the remanent magnetization of the magnetic material.}
      \label{Fig.Halbach_cylinder_drawing}
\end{figure}

The design known as a Halbach cylinder is especially good at producing this type of
magnetic field. A Halbach cylinder is a long cylinder made of a magnetic material with a
bore along the cylinder symmetry axis. The Halbach cylinder can be characterized by
three parameters; the internal and external radii, $r_{\inte}$ and $r_{\exte}$
respectively, and the length, $L$. The magnetic material around the bore is magnetized
such that the direction of magnetization at any point is at an angle
\begin{eqnarray}\label{Eq.Halbach.angle}
\eta{}=2\theta{}
\end{eqnarray}
from the vertical axis \cite{Mallinson_1973,Halbach_1980}. This arrangement means that a
uniform field will be created across the bore in the vertical direction without
creating, in the ideal case, any stray field outside the cylinder. Fig.
\ref{Fig.Halbach_cylinder_drawing} shows a drawing of a Halbach cylinder.

It is well known that the flux density inside the bore of an infinitely long Halbach
cylinder is \cite{Halbach_1980}
\begin{eqnarray}\label{Eq.Halbach.analytic}
B = B_{r}\textrm{ln}\left(\frac{r_{\exte}}{r_{\inte}}\right),
\end{eqnarray}
where $B_{r}$ is the remanent flux density of the magnetic material.

Halbach cylinders have previously been investigated in detail in two dimensions, but
there exist only a few investigations of Halbach cylinders in three dimensions, where
the effect of the finite length of the Halbach cylinder has been studied. In one example
\cite{Mhiochain_1999} the reduction in flux density due to a finite length Halbach
cylinder was investigated, but the field was only calculated for a single fixed length.
An analytical formula for the magnetic flux of a Halbach cylinder of any given length
was derived, however this formula is extremely complicated, making it impractical for
direct application. In another publication \cite{Xu_2004} the effects of a finite length
Halbach cylinder were also explored, as well as the effect of dividing the Halbach
cylinder into a number of segments, each with its own direction of magnetization.
However, both investigations were only performed for one specific Halbach cylinder of a
fixed length.

In this paper the three dimensional Halbach cylinder will be investigated in greater
detail, and the flux density will be computed for a multitude of different
configurations and not only a single specific case.

First, the effect of dividing the Halbach cylinder into segments each with their own
direction of magnetization will be investigated. To measure only the effect of
segmentation the calculations are performed in two dimensions, so that any effects from
a finite length Halbach cylinder are avoided. We then assume that the effect of
segmentation in two dimensions is similar in three dimensions.

Thereafter the Halbach cylinder will be investigated in three dimensions, with focus on
how to build a Halbach cylinder with a certain mean flux density using a minimum of
magnetic material, i.e. find the configuration of $r_{\inte}$, $r_{\exte}$ and $L$ that
generates the strongest flux density for the minimum amount of magnetic material.

Finally it will be investigated if the magnetic flux density  can be improved by placing
additional blocks of permanent magnets on the end faces of the Halbach cylinder.

% It will also be taken into consideration if this additional magnetic material might as
% well be spent on enlarging the external radius of the Halbach cylinder instead.

% if placing additional blocks of permanent magnets, of a certain design and direction of
% magnetization, on the sides of the Halbach cylinder will in any way improve the magnetic
% flux density in and around the cylinder bore, also taking into consideration that this
% material might be spend on enlarging the external radius of the Halbach cylinder
% instead.

The results of this investigation of Halbach cylinder design are useful in many
different fields, e.g. magnetic cooling \cite{Coey_2002} or tabletop NMR
\cite{Moresi_2003}. These applications typically require a flux density of around $1-3$
T, and this is also the range of flux density that we will concern ourselves with in
this paper.

All numerical work in this paper was done using the commercially available finite
element multiphysics program, \emph{Comsol Multiphysics}\cite{Comsol}. The Comsol
Multiphysics code has previously been validated through a number of NAFEMS (National
Agency for Finite Element Methods and Standards) benchmark studies \cite{Comsol_2005}.
%--- Removed the Comsol footnote and placed it in references section

The equation solved in the following simulations is the magnetic vector potential
equation,
\begin{eqnarray}
\nabla{}\times{}(\mu{}_{0}^{-1}\mu{}_{r}^{-1}(\nabla{}\times{}\mathbf{A}-\mathbf{B}_{r}))=0,
\end{eqnarray}
where $\mathbf{A}$ is the magnetic vector potential, $\mathbf{B}_{r}$ is the remanent
flux density, $\mu{}_{0}$ is the permeability of free space and $\mu{}_{r}$ is the relative permeability assumed to be isotropic.

The solver used to solve this equation on the simulation mesh is \emph{Pardiso} which is
a parallel sparse direct linear solver \cite{Schenk_2001,Schenk_2002}.

Boundary conditions are chosen such that the boundaries of the computational volume,
which is many times larger than the Halbach cylinder, are magnetically insulating, while
all other (internal) boundaries are continuous.

%-------------------------------------------------------------------------------------------------------------
%-------------------------------------------------------------------------------------------------------------
%-------------------------------------------------------------------------------------------------------------

\section{Segmented Halbach cylinder}\label{sec.SegmentedHalbach}
An infinitely long Halbach cylinder is equivalent to a two dimensional situation so it
fulfills Eq. \ref{Eq.Halbach.analytic}, if the direction of magnetization varies
continuously through the magnetic material as prescribed by Eq. \ref{Eq.Halbach.angle}.
This continuous variation of the direction of magnetization is often not attainable in
real-world assemblies, and therefore the Halbach cylinder is often made up of segments,
each of which has a direction of magnetization equal to the direction of magnetization
of a continuous Halbach cylinder at the center of the segment.

A Halbach cylinder consisting of $n$ such segments will have its flux density reduced to
\cite{Halbach_1980}
\begin{eqnarray}\label{Eq.Segments_formula}
B(n) = B(\infty{})\frac{\textrm{sin}(2\pi{}/n)}{2\pi{}/n},
\end{eqnarray}
where $B(\infty{})$ is the flux density given by Eq. \ref{Eq.Halbach.analytic}, i.e.
with a continuous magnetization.

We have analyzed the consequence of this segmentation of the Halbach cylinder by computing the mean value of the magnetic flux density inside the Halbach
cylinder bore for a Halbach cylinder consisting of 4, 6, 8, 12, 16, 24 and 32 segments.
The calculations were performed both for a Halbach cylinder consisting of ``perfect''
magnets, i.e. with a relative permeability, $\mu{}_{r}$, of 1, and magnets where actual
material properties were taken into account by increasing $\mu{}_{r}$ to $1.05$. The
magnetic material was assumed to have a remanent flux density of 1.4 T, equal to
standard grade N48 Neodymium-Iron-Boron (NdFeB) magnets \cite{Standard}. This value for
the remanent flux density will be used in the remainder of this paper.
%--- Removed footnote and included it as a sentence

\begin{figure}[!t]
  \centering
   \includegraphics[width=\columnwidth]{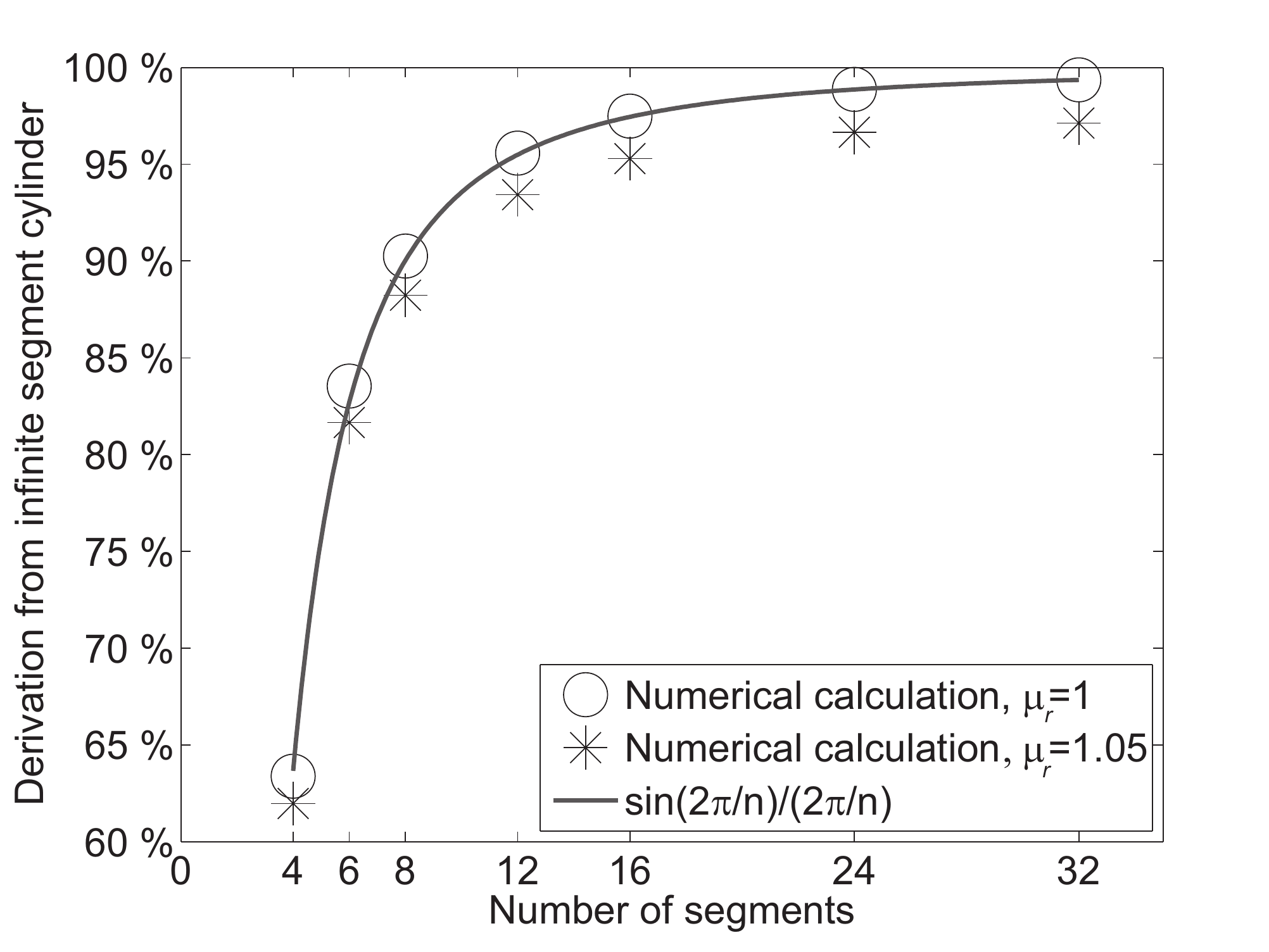}
      \caption{Dividing a Halbach cylinder into $n$ segments makes the flux density differ from that of a
       perfect Halbach cylinder. The deviation is given by Eq. \ref{Eq.Segments_formula}
       for "perfect magnets", $\mu{}_{r}=1$, while magnets with $\mu{}_{r}=1.05$ deviate more from the ideal Halbach cylinder.}
       \label{Fig.Number_of_segments_vs_derivation_from_infinite_cylinder}
\end{figure}

The results of the computations together with Eq. \ref{Eq.Segments_formula} are shown in
Fig. \ref{Fig.Number_of_segments_vs_derivation_from_infinite_cylinder}. Here it is seen
that Eq. \ref{Eq.Segments_formula} describes the numerical data with $\mu_{r}=1$ extremely
well. It is also seen that choosing a small number of segments severely limits the flux
density.

Based on the results shown in Fig.
\ref{Fig.Number_of_segments_vs_derivation_from_infinite_cylinder} we choose, in the
calculations and computations presented in the following sections, to use a 16 segmented
Halbach cylinder with $\mu{}_{r}=1.05$. This configuration obtains 95\% of the flux
density of a perfect Halbach cylinder and is realizable in real-world assemblies.

Having determined the configuration to be used in the following simulations we now
proceed to investigate if there exist optimal dimensions for a Halbach cylinder design.
For this three dimensional simulations must be used, in order to study how the loss of
flux density through the ends of the cylinder bore varies with $r_{\exte}$ and $L$.

%-------------------------------------------------------------------------------------------------------------
%-------------------------------------------------------------------------------------------------------------
%-------------------------------------------------------------------------------------------------------------

\section{Halbach cylinder 3D study}\label{Sec.Halbach-cylinder-parameter-survey}
\begin{figure}[!t]
  \centering
   \includegraphics[width=\columnwidth]{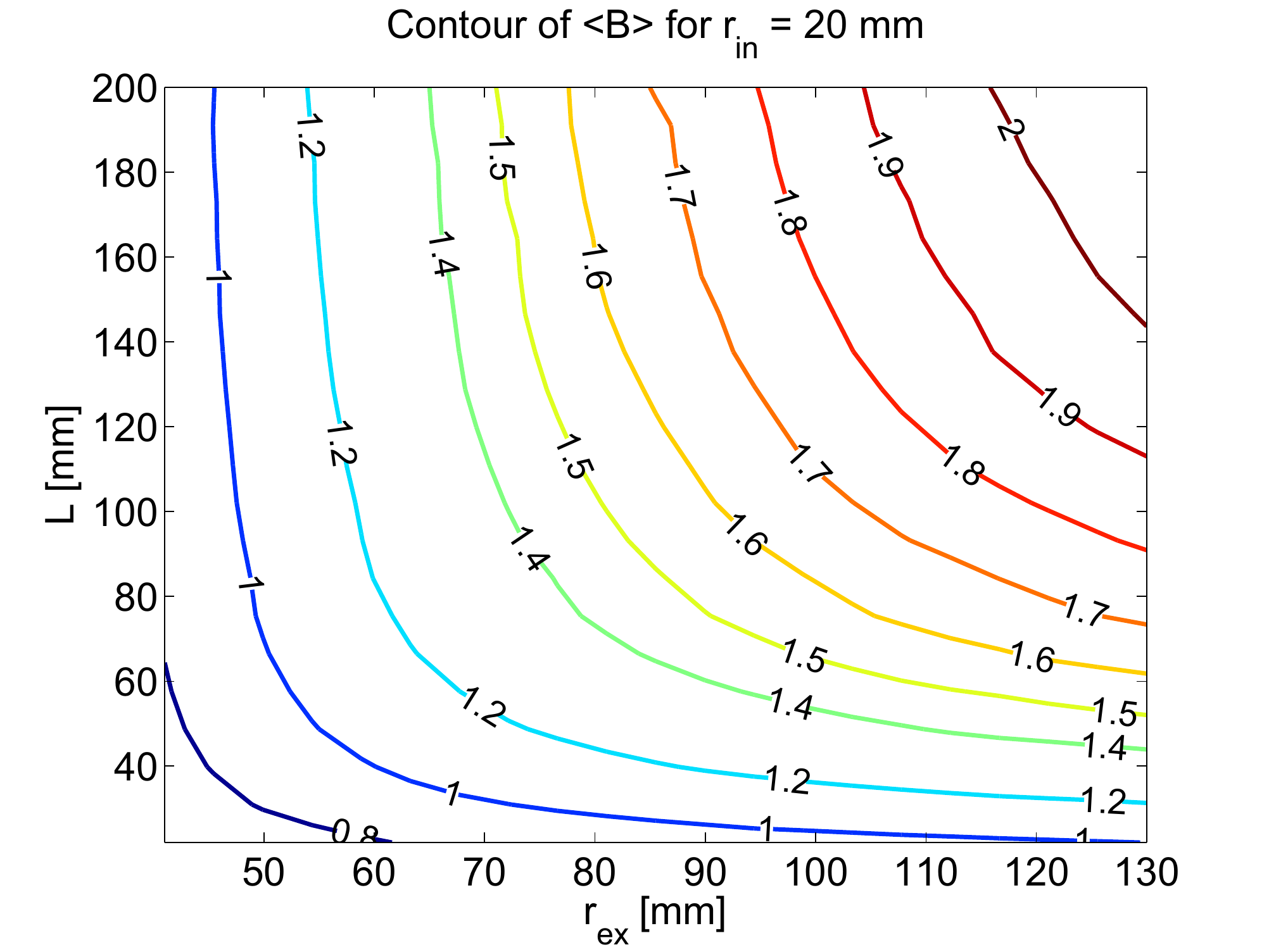}
      \caption{ Contours of the mean flux density for the Halbach cylinders with
         $r_{\inte}=$ 20 mm. Each contour is labeled by its mean flux density. As is expected the maximum flux density is obtained by maximizing
         both $r_{\exte}$ and $L$.} \label{Fig.Contour_Length_vs_Rexternal_vs_B_-_Rinternal_0.02}
\end{figure}

A parameter study of Halbach cylinder configurations have been performed by varying the
parameters $L$, $r_{\exte}$ and $r_{\inte}$ as given in Table
\ref{Table.Halbach-cylinders-parameters}. In each of the $90\times{}90\times{}3$
configurations the mean flux density of the magnetic field inside the cylinder bore have
been computed. The results are shown as a contour plot of the mean flux density as a
function of $L$ and $r_{\exte}$ in Fig.
\ref{Fig.Contour_Length_vs_Rexternal_vs_B_-_Rinternal_0.02} for $r_{\inte}=$ 20 mm.

\begin{table}[!t]
\begin{center}
\caption{The variation of the Halbach parameters. In total there are
$90\times{}90\times{}3$ different
configurations.}\label{Table.Halbach-cylinders-parameters}
\begin {tabular}{l|ccc}
 & From & To & Stepsize \\
 & [mm] & [mm] & [mm] \\ \hline
$L$           & 41  & 130 &  \hspace{0.2cm}1\\
$r_{\exte}$   & 22  & 200 &  \hspace{0.2cm}2\\
$r_{\inte}$   & 10  &  \hspace{0.2cm}30 & 10
\end {tabular}
\end{center}
\end{table}

Fig. \ref{Fig.Contour_Length_vs_Rexternal_vs_B_-_Rinternal_0.02} shows that the
configuration producing the strongest mean flux density is the configuration with the
largest $r_{\exte}$ and $L$. This is in agreement with Eq. \ref{Eq.Halbach.analytic} and
the fact that for a long Halbach cylinder the loss of flux through the ends of the
cylinder will be relatively smaller than for a short cylinder.

\begin{figure}[!t]
  \centering
   \includegraphics[width=0.93\columnwidth]{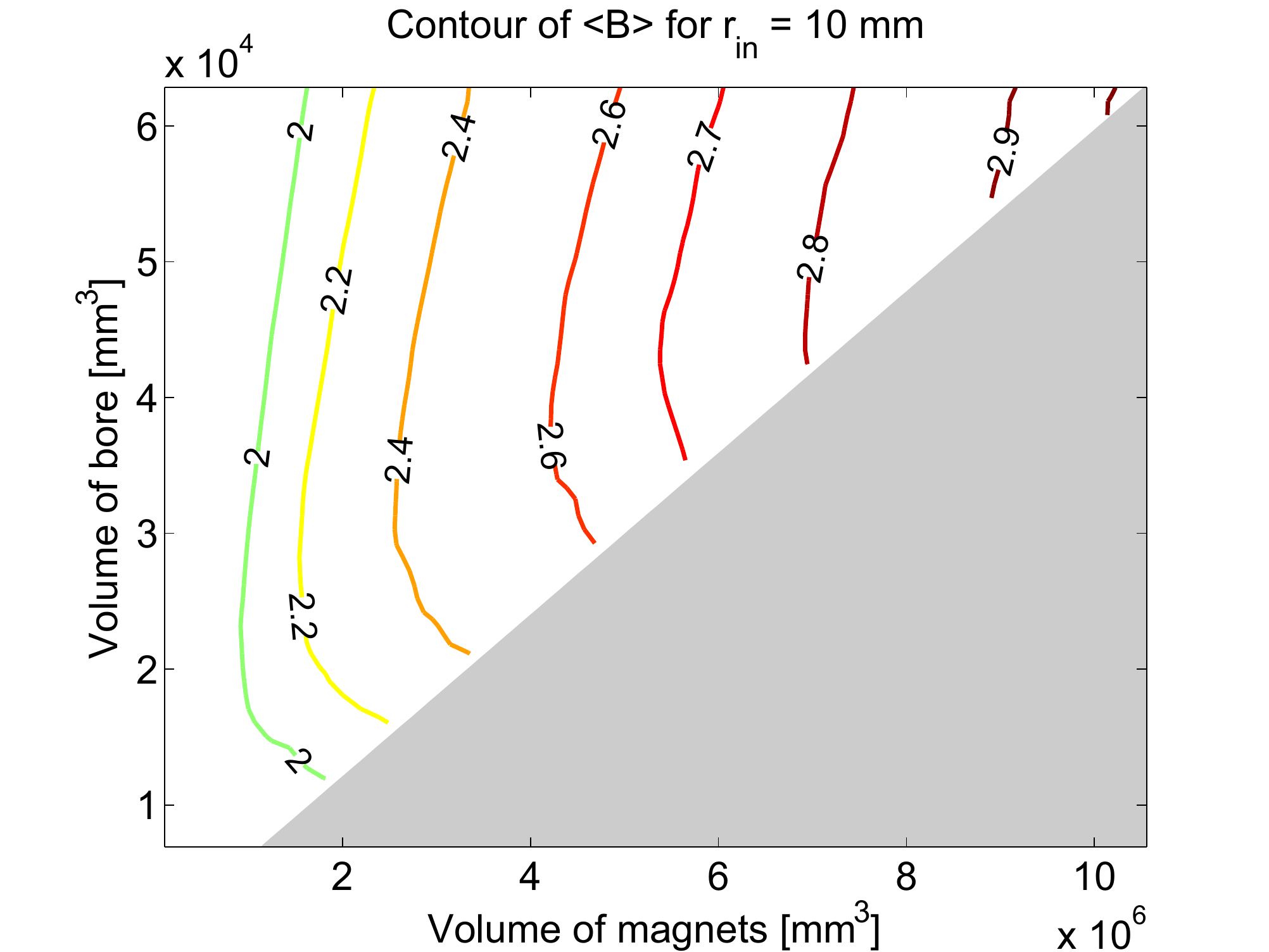}
      \caption{ Contours of the mean flux density as a function of the volume of magnets used and
                the volume of the cylinder bore for $r_{\inte}=$ 10 mm. It is seen that the volume of the
                bore can be significantly increased by slightly increasing the volume of the magnets. Note that the range is not the same on the two
                axes. A look-up table is necessary such that each data point (these are not shown)
                is uniquely tied to a specific Halbach cylinder, i.e. a given $r_{\inte}$, $r_{\exte}$ and $L$.}
\label{Fig.Vmag-Vhole-B-r-0.01-contour}
\end{figure}

\begin{figure}[!t]
  \centering
   \includegraphics[width=\columnwidth]{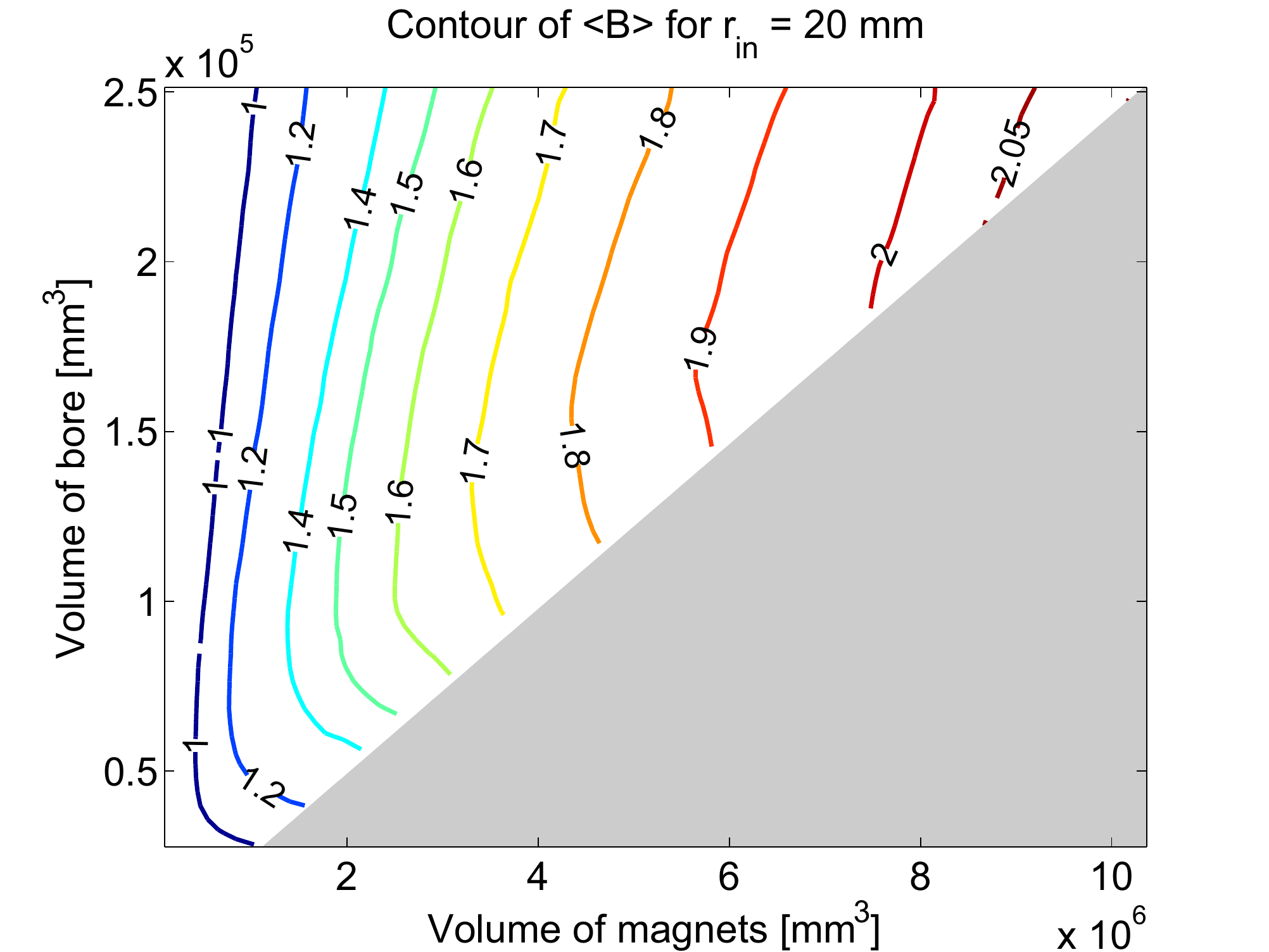}
      \caption{ Contours of the mean flux density as a function of the volume of magnets used and
                the volume of the cylinder bore for $r_{\inte}=$ 20 mm. The conclusion of Fig. \ref{Fig.Vmag-Vhole-B-r-0.01-contour} applies here as well.} \label{Fig.Vmag-Vhole-B-r-0.02-contour}
\end{figure}

\begin{figure}[!t]
  \centering
   \includegraphics[width=\columnwidth]{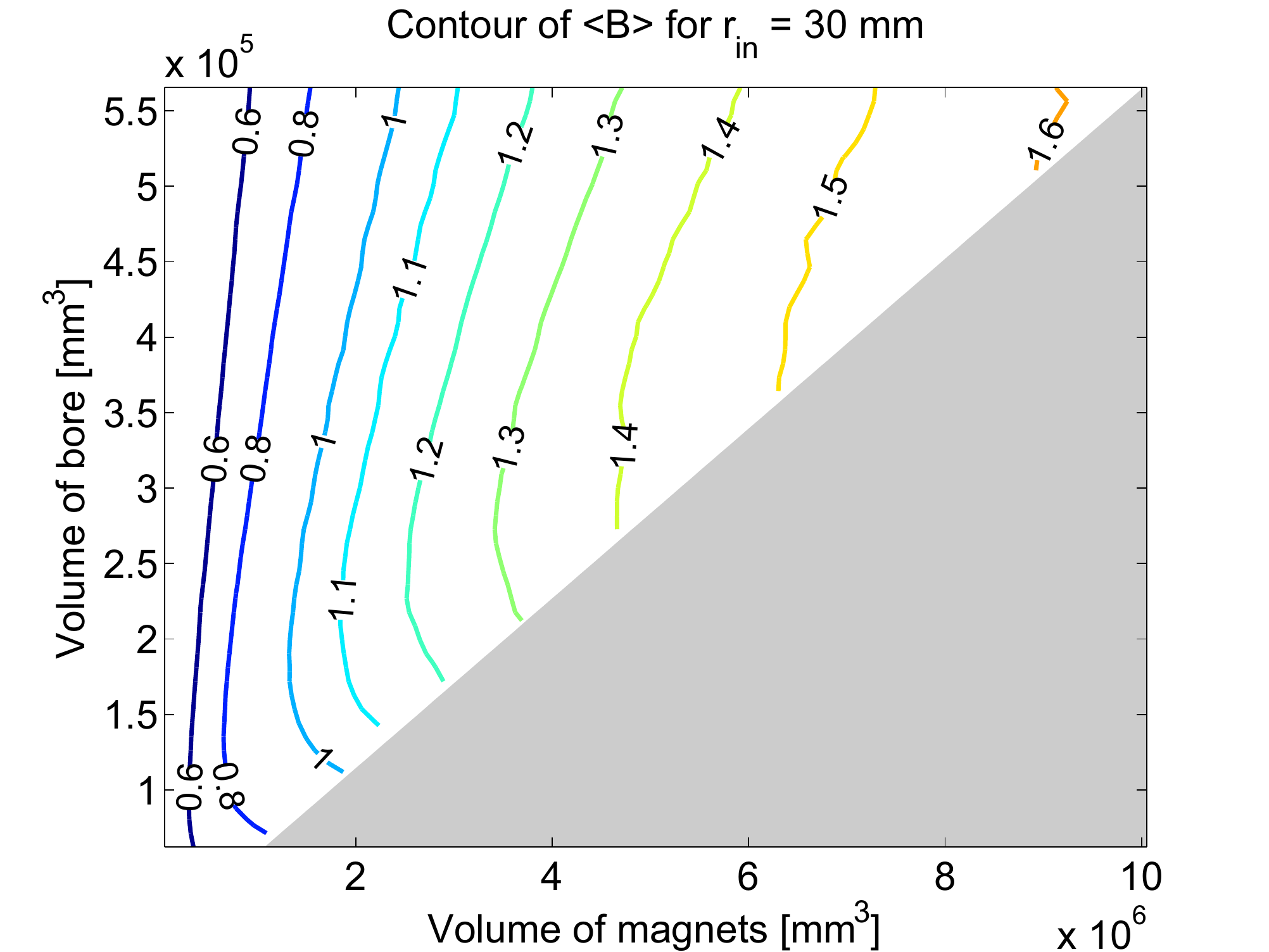}
      \caption{ Contours of the mean flux density as a function of the volume of magnets used and
                the volume of the cylinder bore for $r_{\inte}=$ 30 mm. The conclusion of Fig. \ref{Fig.Vmag-Vhole-B-r-0.01-contour} applies here as well.} \label{Fig.Vmag-Vhole-B-r-0.03-contour}
\end{figure}

It is not sufficient to characterize a design only by the value of the mean flux
density. It should be considered that increasing the length of the Halbach cylinder
increases the volume of the bore, thus allowing a larger sample to be placed inside the
Halbach cylinder bore. On the other hand increasing the external radius does not affect the volume of the bore. Consequently a better way of characterizing each
Halbach cylinder configuration is by the volume of its magnets and the volume of the
bore, and then calculating contour plots with lines of equal mean flux density. These
are shown in Figs.
\ref{Fig.Vmag-Vhole-B-r-0.01-contour}-\ref{Fig.Vmag-Vhole-B-r-0.03-contour} for the
three different values of $r_{\inte}$. On Figs.
\ref{Fig.Vmag-Vhole-B-r-0.01-contour}-\ref{Fig.Vmag-Vhole-B-r-0.03-contour} the volume
of the bore scales directly with the length of the Halbach cylinder because the internal
radius is kept constant in each figure.


Looking at, e.g., Fig. \ref{Fig.Vmag-Vhole-B-r-0.02-contour} it can be seen that for a
mean flux density of 1.6 T a Halbach cylinder can be constructed with a $\sim$50\%
increase in magnetic material but a $\sim$250\% larger volume of the bore compared to
the design using the least amount of magnetic material.

It is possible to attain this substantial increase in the volume of the bore because the
latter configuration is a very long Halbach cylinder with a small external radius, while
the configuration with the smallest volume of the magnets is a short Halbach cylinder
with a large external radius. In these two configurations the shape of the bore is
different, but the mean flux density is the same.

\begin{figure}[!t]
  \centering
   \includegraphics[width=\columnwidth]{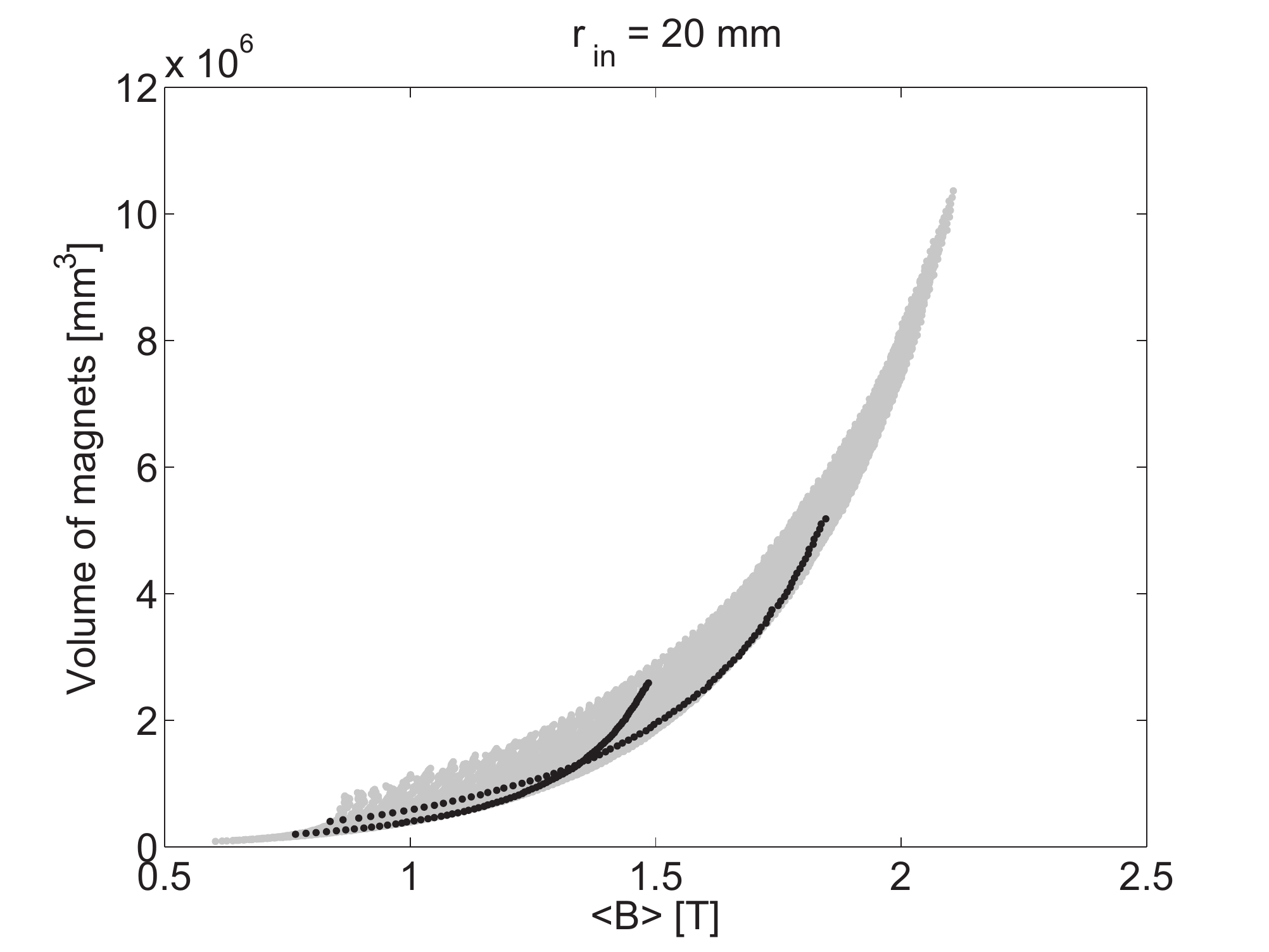}
      \caption{The total volume of the magnetic material as a function of the mean flux density for
$r_{\inte}=$ 20 mm. The data points somewhat resemble a feather, and so this plot will
be referred to as the ``feather plot''. The data have been produced in series where
$r_{\exte}$ has been fixed and $L$ has been varied. Two of these data series have been
highlighted in black and starting from the left both series can be seen to first
approach the rightwards edge of the ``feather'' and then leave it again. Similar plots
exist for $r_{\inte}=10$ mm and 30 mm.} \label{Fig.B-TotalVolume-r-in-0-02}
\end{figure}

In Fig. \ref{Fig.B-TotalVolume-r-in-0-02} the total volume of the magnetic material is
shown as a function of the mean flux density in the bore for $r_{\inte}=20$ mm. In this
plot there are $90\times{}90$ data points. Two data series where $r_{\exte}$ has been
fixed and $L$ has been varied are highlighted on the plot (one could also have chosen to
vary $r_{\exte}$ and kept $L$ fixed. The curves look the same). Here one can see that as
$L$ is increased the mean flux density is increased as well. At some point each
data series becomes the rightwards edge of the ``feather'', and then the increase in the
volume of the magnets with mean flux density becomes too steep and the data points move
upwards, leaving the edge of the ``feather''.

The reason for this behavior is that the data series start with a short Halbach
cylinder. This configuration loses a lot of flux through the ends of the cylinder and so
as the length is increased the average flux density increases quite rapidly. When a
certain length of the Halbach cylinder is reached there is not as much to be gained by
increasing the length of the cylinder further and so the average flux density only
increases slowly as the volume of the magnets is increased.

As this is the case for all data series where the length of the Halbach cylinder is
gradually increased it is possible to characterize the rightwards edge of the
``feather'' as the optimal configuration, i.e. the configuration with the smallest
volume of the magnets at a given mean flux density.

\begin{figure}[!t]
  \centering
   \includegraphics[width=\columnwidth]{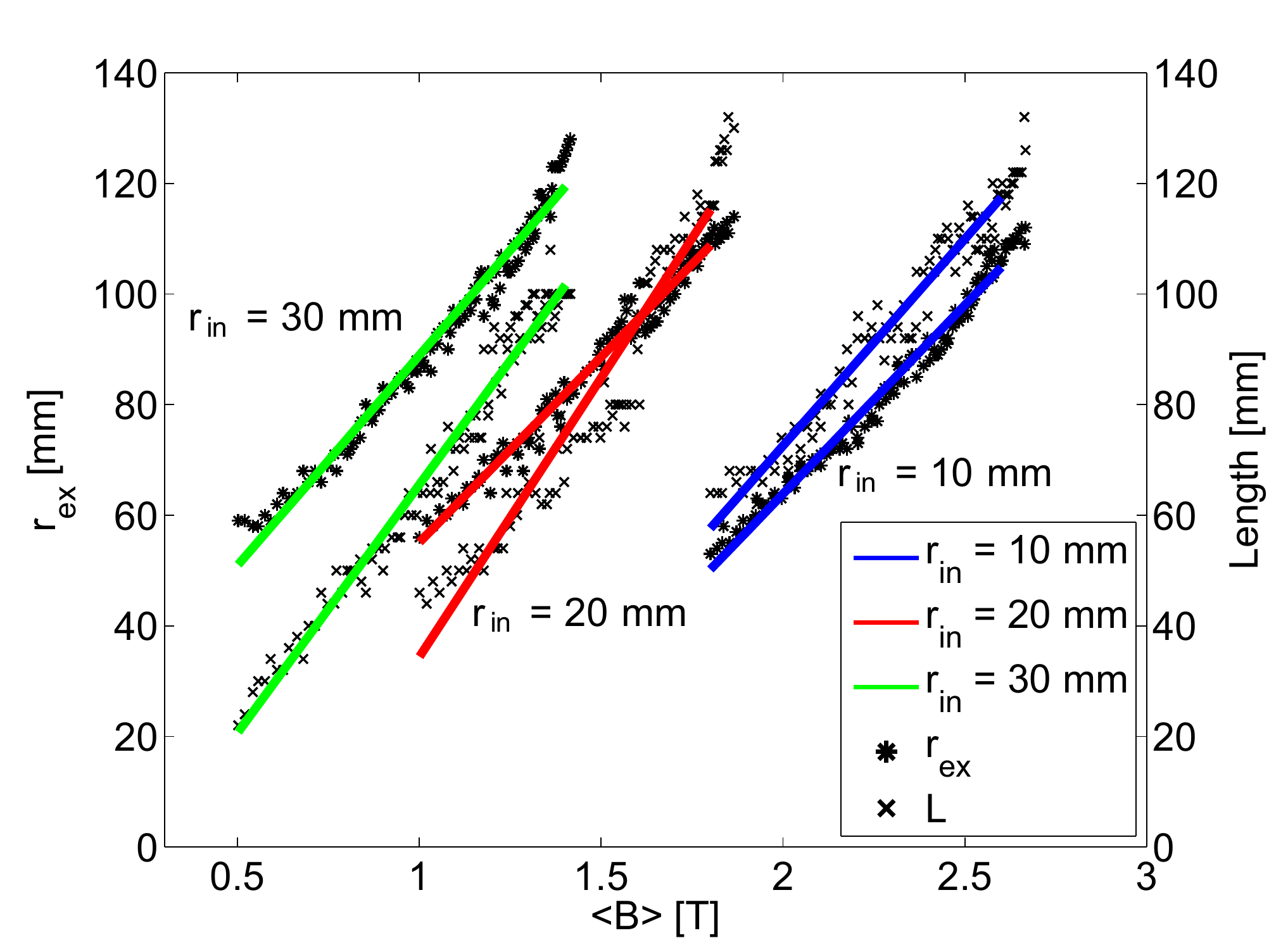}
      \caption{ The optimal $r_{\exte}$ and $L$ as functions of the mean flux density in the
    cylinder bore for Halbach cylinders with $r_{\inte}=$ 10, 20 and 30 mm. Building a Halbach
    cylinder with dimensions different from the dimensions given here means that more magnetic material
    is used than need be, if one is only interested in obtaining the maximum flux density possible
    and does not care about the size of the cylinder bore.}
\label{Fig.B-r-external-r-in-all}
\end{figure}

In Fig. \ref{Fig.B-r-external-r-in-all} the values of $r_{\exte}$ and $L$ are plotted as
functions of the mean flux density for the optimal points. Thus one can directly use
this figure to find the external radius and length for the Halbach cylinder with the
minimum volume of magnets at a given mean flux density. Straight lines have been fitted
to the data.

The conclusion of this parameter investigation is twofold. First, it can be concluded
from Figs. \ref{Fig.Vmag-Vhole-B-r-0.01-contour}-\ref{Fig.Vmag-Vhole-B-r-0.03-contour}
that it is possible, at a constant mean flux density, to increase the volume of the bore
significantly by only increasing the volume of the magnets slightly for a Halbach
cylinder with a fixed $r_{\inte}$. Secondly, the length and external radius of the
minimum magnetic material Halbach cylinder at a given mean flux density was found and
can be read off directly from Fig. \ref{Fig.B-r-external-r-in-all}. Thus if one wishes
to build a Halbach cylinder with a given mean flux density and the limiting factor is
the price of magnetic material, one should always choose the optimal configuration from
this figure.

A few remarks on the precision of the simulations are in order. With the chosen boundary
conditions, i.e. a magnetically insulating computational volume, it is important that
the computational volume is large enough that the insulating boundaries do not effect
the calculations. Also, as the solution method used is a finite element method, the mesh
applied to the geometry must be as detailed as needed for the desired precision. The
resolution of the mesh used for the simulations presented in this paper are chosen such
that the results have a high degree of precision. To give an example, the mean flux
density of the cylinder bore was calculated at different precisions for a random Halbach
cylinder. The Halbach cylinder chosen had $r_{\inte}=20$ mm, $r_{\exte}=102$ mm and
$L=70$ mm, and a mean flux density of 1.54 T. This result, calculated using the
precision used throughout this paper, deviated by only 1.13\% from a simulation using
173\% more mesh elements all in all, and 1845\% more mesh elements in the cylinder bore.
The influence of the size of the computational volume on the mean flux density in the
bore has also been tested for a number of different values of $r_{\inte}$, $r_{\exte}$
and $L$ and found to be less than 1\%. Thus we conclude that at least the relative
precision of the numerical experiments is satisfactory.

Although the above results are useful in choosing the optimal Halbach cylinder design,
alternative methods for improving the design of a Halbach cylinder exists. The problem
with especially the short Halbach cylinders is that they lose a relatively large amount
of flux through the ends of the cylinder. This is the reason that their flux density is
not well described by Eq. \ref{Eq.Halbach.analytic}. In the next section it is
investigated if it is possible to limit the amount of escaping flux through the ends of
the cylinder by appending blocks of permanent magnets to the end faces of the Halbach
cylinder and thus in this way improve the design.

%-------------------------------------------------------------------------------------------------------------
%-------------------------------------------------------------------------------------------------------------
%-------------------------------------------------------------------------------------------------------------

\section{Improving the Halbach cylinder design}\label{ref.ImprovingHalbachdesign}
The main loss of flux from the bore of the Halbach cylinder is through the ends of the
cylinder bore. It has previously been shown \cite{Potenziani_1987} that by ``covering''
the ends of the Halbach cylinder with magnetic blocks in the shape of an equipotential
surface, all of the flux can be confined inside the Halbach cylinder. However this also
blocks access to the cylinder bore. Instead we propose that some of the escaping flux
may be confined by placing additional magnets, of a given size and direction of
magnetization, at the end faces of the cylinder, in such a way that they do not block
access to the cylinder bore but still increase the flux density in the cylinder bore and
ensure a low flux density outside of the Halbach cylinder.

In this section we investigate what specific design and placement is optimal for these
additional magnets. We also discuss whether it is better to use the additional magnets
or if one might as well use the additional magnetic material for building a larger
Halbach cylinder.

\begin{figure}[!t]
  \centering
   \includegraphics[width=1\columnwidth]{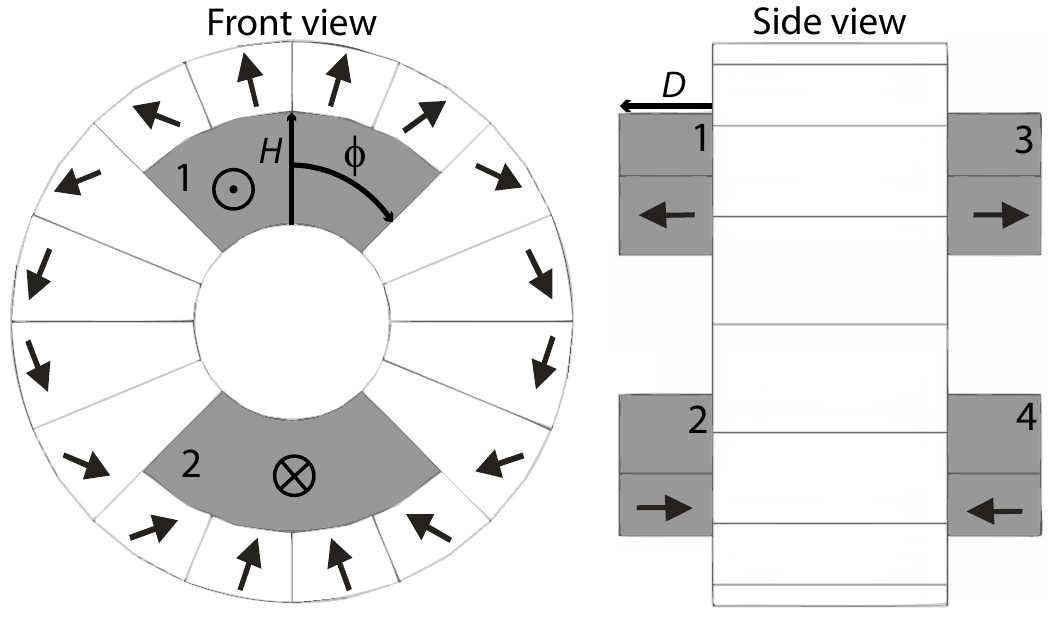}
      \caption{This figure shows the height, $H$, angular span, $\phi{}$, and depth, $D$, of the additional round blocks, colored in gray.
      The blocks are always symmetrically placed. The black arrows shows the direction of magnetization.
      The additional blocks diagonally opposite each other have the same direction of magnetization.}
      \label{Fig.Halbach_extra_blocks_design}
\end{figure}

To maximize the amount of magnetic material capable of being used in the additional
blocks we use a design of the additional blocks that follows the curvature of the
cylinder bore, i.e. a circular design as can be seen in Fig.
\ref{Fig.Halbach_extra_blocks_design}. In total four additional blocks are used, placed
symmetrically around the Halbach cylinder symmetry axis. In this design an additional
block is characterized by three parameters, namely the angular extent of a block,
$\phi{}$, the block's depth, $D$, and height, $H$. The direction of magnetization of the
individual additional block is perpendicular to the Halbach cylinder end face.
Furthermore the blocks diagonally opposite have the same direction of magnetization.

The height, $H$, angular span, $\phi{}$, and depth, $D$, of the additional blocks are
varied to find the optimal configuration for several different Halbach cylinders.
Calculating the flux density for each of the original $90\times{}90\times{}3$ Halbach
cylinders with additional magnets is a too time consuming task, and thus the
calculations were only done on a few carefully chosen Halbach cylinder designs. These
are given in Table \ref{Table.Halbach-cylinders-with-extra-blocks}.

\begin{table}[!t]
\begin{center}
\caption{The additional magnets were placed on the four different Halbach cylinders
given in column 5. The parameters of the additional blocks were varied as given in this
table resulting in $10\times{}10\times{}8$ different configurations of the additional
blocks for each Halbach cylinder.}
\begin {tabular}{l|rrr||r}
 & From & To & Step & Halbach  dim-  \\
 &      &    & size         & ensions [mm]\\ \hline
%$\phi{}$  \hspace{0.104cm}[$^{\circ}$]   & $10$   & $80$   &  $10$        & $L=\hspace{0.2cm}50$ mm\\
%$H$ [mm]                         & 60     & 25     & 5            & $r_{\exte}=\hspace{0.2cm}60$ mm\\
%$D$  \hspace{0.04cm}[mm]           & 25     & 2.5    & 2.25         & $r_{\inte}=\hspace{0.2cm}20$ mm \\ \hline
$\phi{}$  \hspace{0.104cm}[$^{\circ}$]  & $10$   & $80$   &  $10$        & $L=100$ \\
$H$ [mm]                           & 100    & 30     & 10            & $r_{\exte}=100$ \\
$D$  \hspace{0.04cm}[mm]           & 50     & 5      & 5             &
$r_{\inte}=\hspace{0.16cm}20$  \\ \hline
$\phi{}$  \hspace{0.104cm}[$^{\circ}$]  & $10$   & $80$   &  $10$    & $L=\hspace{0.16cm}92$ \\
$H$ [mm]                           & 130    & 25     & 15           & $r_{\exte}=130$ \\
$D$  \hspace{0.04cm}[mm]           & 46     & 4.6    & 4.6          &
$r_{\inte}=\hspace{0.16cm}20$ \\ \hline
$\phi{}$   \hspace{0.104cm}[$^{\circ}$]  & $10$   & $80$   &  $10$        & $L=200$ \\
$H$ [mm]                           & 60     & 25     & 5            & $r_{\exte}=\hspace{0.16cm}60$ \\
$D$  \hspace{0.04cm}[mm]           & 100    & 10     & 10           &
$r_{\inte}=\hspace{0.16cm}20$ \\ \hline
$\phi{}$   \hspace{0.104cm}[$^{\circ}$]  & $10$   & $80$   &  $10$        & $L=\hspace{0.16cm}50$ \\
$H$ [mm]                           & 130    & 25     & 15           & $r_{\exte}=130$ \\
$D$  \hspace{0.04cm}[mm]           & 25     & 2.5    & 2.25         &
$r_{\inte}=\hspace{0.16cm}20$  \\ \hline
\end {tabular}
\label{Table.Halbach-cylinders-with-extra-blocks}
\end{center}
\end{table}

% In these simulations the parameters of the additional blocks were varied as given in
% Table \ref{Table.Halbach-cylinders-with-extra-blocks}. The heights of the additional
% blocks were varied from the external radius of the Halbach cylinder to 25 mm in seven
% equidistant steps and the depths of the blocks were varied from a depth equal to half
% the length of the Halbach cylinder to a tenth of this in ten equidistant steps.

\begin{figure}[!t]
  \centering
   \includegraphics[width=\columnwidth]{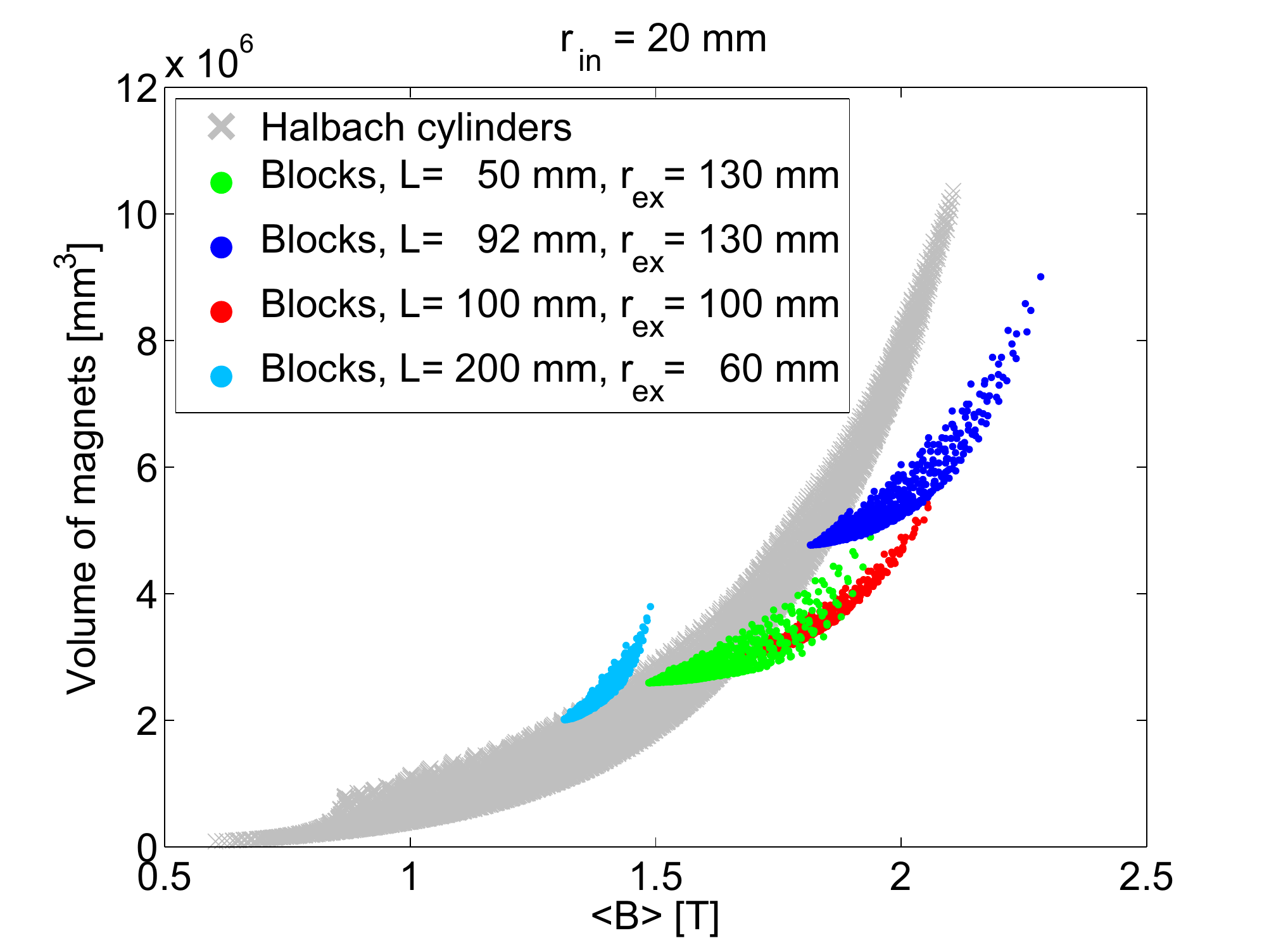}
      \caption{ Placing additional blocks on a specific Halbach cylinder
improves the mean flux density in the cylinder bore but the
improvement depends on $L$ and $r_{\exte}$ of the Halbach cylinder. The legend shows
which Halbach cylinder the additional blocks was placed on. Without the additional
blocks the figure is identical to Fig. \ref{Fig.B-TotalVolume-r-in-0-02}.}
\label{Fig.B-TotalVolume-r-0.02-extra-blocks}
\end{figure}

The results of the simulations are shown in Fig.
\ref{Fig.B-TotalVolume-r-0.02-extra-blocks}. Here the mean flux density in the bore as a
function of the total volume of the magnetic material used in the simulated design is
shown. The figure shows both the Halbach cylinders without any additional blocks, and
the simulations of the Halbach cylinders with additional blocks.

One can see from the figure that placing additional blocks on a relatively short Halbach
cylinder will improve the mean flux density in the cylinder bore significantly.

%-------------------------------------------------------------------------------------------------------------
%-------------------------------------------------------------------------------------------------------------
%-------------------------------------------------------------------------------------------------------------
\section{Halbach cylinders for use in magnetic cooling}\label{ref.Halbachcylinderinmagneticrefrigeration}

We have shown that using additional blocks of magnets on the sides of the Halbach
cylinder can increase the mean flux density in the cylinder bore. However in some cases
the additional magnetic material might as well be used to enlarge the Halbach cylinders
external radius and in this way also increase the flux density.
% Although one can see
% from Fig. \ref{Fig.B-TotalVolume-r-0.02-extra-blocks} that the blocks in some cases are
% clearly superior to simply expanding the external radius this might not be the case for
% all configurations.
We will consider this more closely in the context of one particular application for
Halbach cylinders, namely magnetic cooling. For this type of application the Halbach
cylinder must be designed such that it has a high flux density in a large volume and
with a minimum of magnetic material.

The magnetic cooling process relies on a magnetocaloric material. The temperature of
such a material is increased upon the application of a magnetic field and decreased
again upon the removal of the magnetic field. A large number of different materials have
been suggested as the active component of a magnetic refrigeration machine
\cite{Gschneidner_2005}.

From experimental studies it is known that the adiabatic temperature change of
Gadolinium, the ``benchmark'' magnetocaloric material at room temperature, has a
magnetocaloric effect that scales with the flux density of the magnetic field
\citep{Pecharsky_2006} to the power of $0.7$ . This is in good accordance with the power
of $\frac{2}{3}$ predicted by mean field theory \cite{Osterreicher_1984}.

However, it is not only the flux density inside the cylinder bore that is of importance
to the magnetocaloric effect. The volume outside the cylinder bore where the
magnetocaloric material is placed when it is moved to the ``out of field'' position is
also important. In order to maximise the magnetocaloric effect the flux density in this
region must be as low as possible. It can of course be argued that one could simply move
the magnetocaloric material further away than right outside the end of the cylinder
bore, but this would increase the physical size of the magnetic refrigeration machine.
Finally, it is important that the cylinder bore has as large a volume as possible and
that the volume of the magnets be as small as possible.

\begin{figure}[!t]
   \centering
   \includegraphics[width=.7\columnwidth]{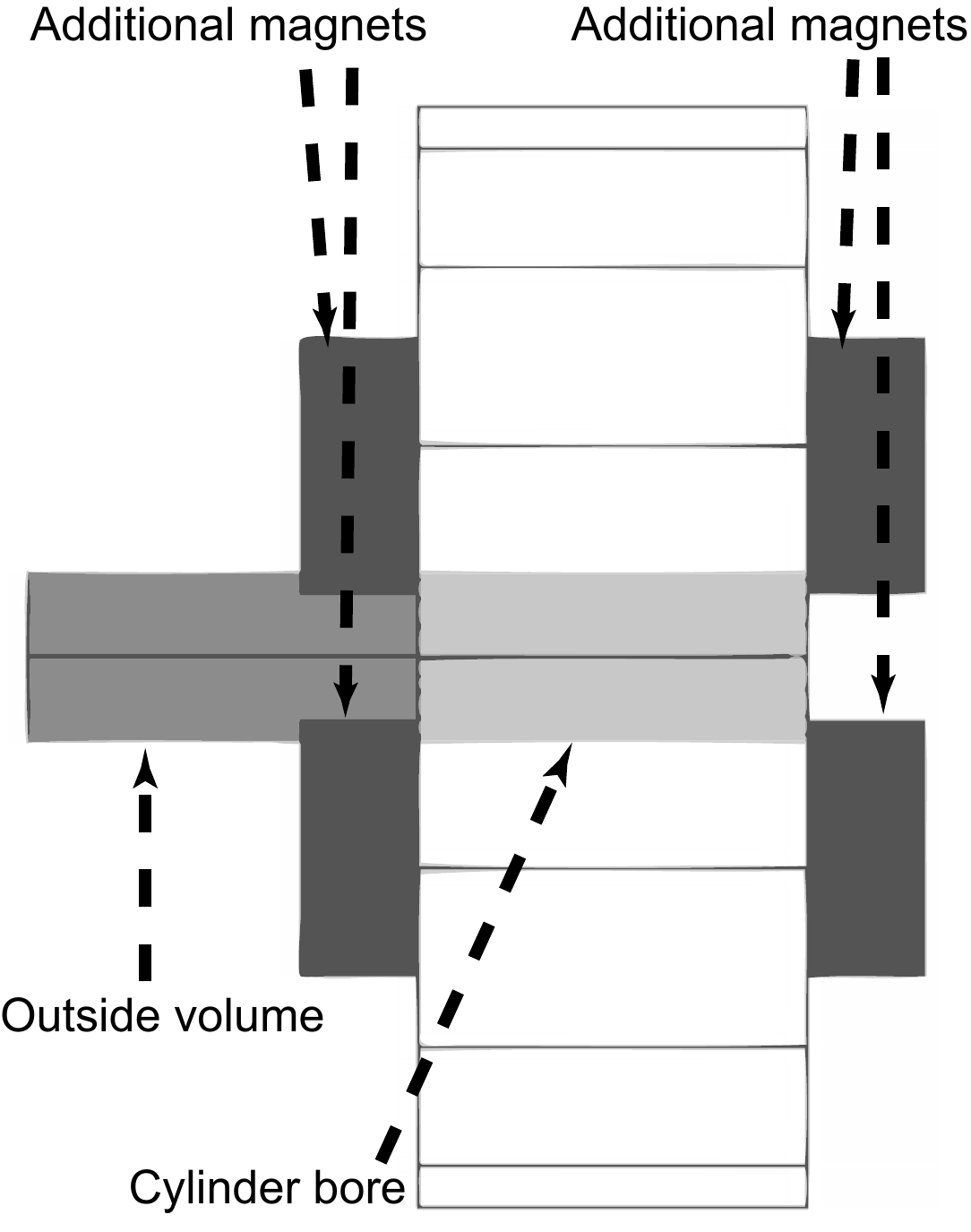}
      \caption{A side view of the Halbach cylinder with additional blocks.
The volume inside which $\avetxt{0.7}{out}$ is calculated is shown. The volume is
identical in shape to the cylinder bore, also shown, and is located directly outside the
end of the cylinder bore. Also shown are the additional blocks of magnets. The edges of
the 16 segments that make up the Halbach cylinder can also be seen on the figure.}
\label{Fig.Show_integration_volumes}
\end{figure}

Taking all this into account we propose to characterize a configuration of magnets for
use in magnetic cooling applications by the parameter
\begin{eqnarray}\label{Eq.Magcool-parameter}
\Lambda_{\mathrm{cool}} \equiv (\avenotxt{0.7}-\avetxt{0.7}{out})\frac{V_{\mathrm{field}}}{V_{\mathrm{mag}}}\;P_{\mathrm{field}},
\end{eqnarray}
where $V_{\mathrm{mag}}$ is the volume of the magnets, $V_{\mathrm{field}}$ is the
volume with a high flux density, i.e. the volume of the cylinder bore,
$P_{\mathrm{field}}$ is the fraction of the total volume of the cylinder bore and the
volume outside the cylinder bore that is filled with magnetocaloric material,
$\avenotxt{0.7}$ is the volume average of the flux density in the high flux volume, i.e.
the cylinder bore, to the power of 0.7 and $\avetxt{0.7}{out}$ is the volume average of
the flux density to the power of 0.7 in the region shown in Fig.
\ref{Fig.Show_integration_volumes}, i.e. the volume just outside the cylinder bore where
the magnetocaloric material is placed when it is moved out of the magnetic field. It has
the same size and shape as the cylinder bore.

\begin{figure}[!t]
  \centering
   \includegraphics[width=\columnwidth]{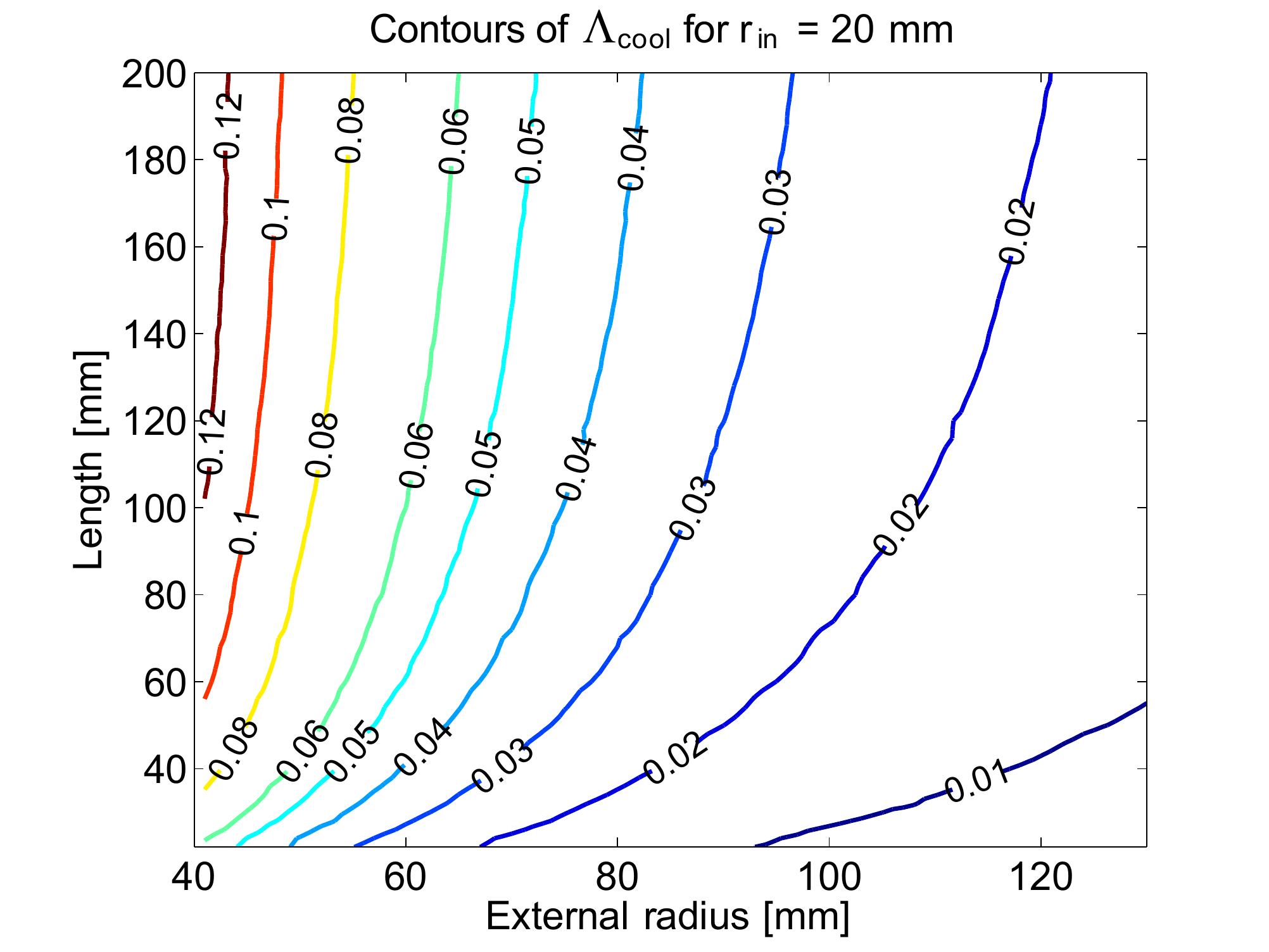}
      \caption{ A contour plot showing the magnetic cooling parameter, $\Lambda_{\mathrm{cool}}$, defined in Eq. \ref{Eq.Magcool-parameter}
for the Halbach cylinders without additional blocks and with $r_{\inte} = 20$ mm. The
optimum design is the longest and thinnest Halbach cylinder.}
\label{Fig.Magcool-parameter-Halbachs}
\end{figure}

The magnetic cooling parameter is shown for the Halbach cylinders without additional
blocks and with $r_{\inte} = 20$ mm in Fig. \ref{Fig.Magcool-parameter-Halbachs} for
$P_{\mathrm{field}}=0.5$ i.e. we assume that the total volume is only half filled with
magnetocaloric material at any given moment. Here we see that the optimal design is the
Halbach cylinder with the largest $L$ and smallest $r_{\exte}$. Note that this design is
not the overall optimal design, as it lies on the edge of the parameter space, i.e.
simulations have not been conducted with a larger $L$ and smaller $r_{\exte}$.

There are several reasons that the long, thin Halbach cylinder has the largest
$\Lambda_{\mathrm{cool}}$. The primary cause is due to the fact that the loss of flux
through the ends of the cylinder bore is greatly reduced in the long Halbach cylinder.
Another effect is due to its long length, the volume inside which $\avetxt{0.7}{out}$ is
calculated is also long (as previously mentioned the shape of the cylinder bore and this
volume are identical), and thus the end furthest from the cylinder bore will only
experience a very small field, thus lowering $\avetxt{0.7}{out}$. For practical
applications one would choose to optimize $\Lambda_{\mathrm{cool}}$ under a criterium of
a minimum flux density in the cylinder bore, i.e. find the Halbach cylinder with the
maximum $\Lambda_{\mathrm{cool}}$ that at the same time has a minimum flux density of
e.g. 1 T in the cylinder bore.

We are also interested in knowing what effect the additional blocks of magnets have on
the magnetic cooling parameter. In Fig. \ref{Fig.B-TotalVolume-r-0.02-extra-blocks} we
saw that the additional blocks increase the flux density in the cylinder bore, but this
might not mean that $\Lambda_{\mathrm{cool}}$ is increased as well, as additional
magnetic material is also used.

\begin{figure}[!t]
  \centering
   \includegraphics[width=\columnwidth]{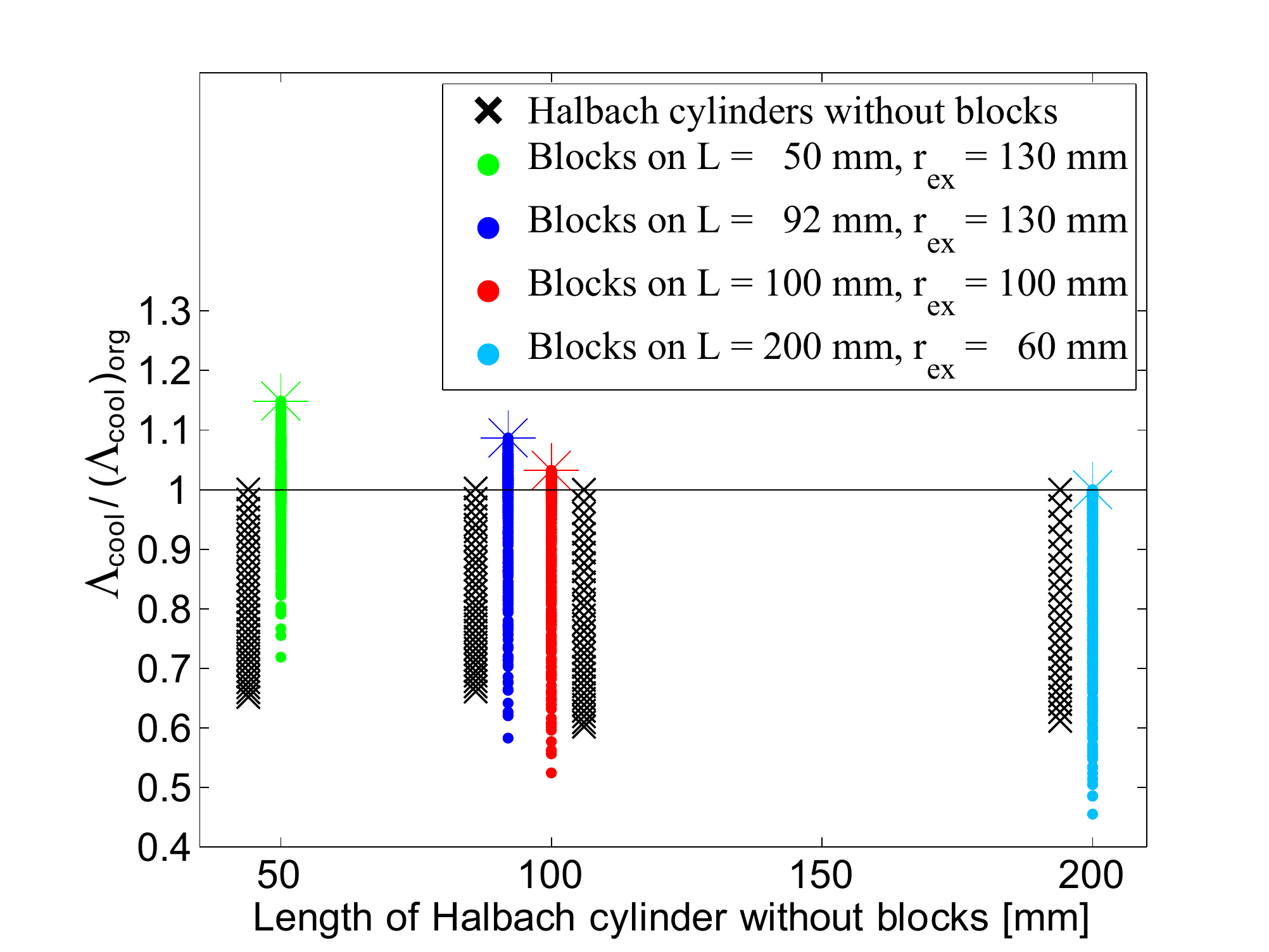}
      \caption{ The relative improvement of the magnetic cooling parameter compared to the original Halbach
cylinder without any additional blocks, for the designs listed in Table
\ref{Table.Halbach-cylinders-with-extra-blocks}. The design most improved is the short
Halbach cylinder with a large external radius, i.e. $L= 50$ mm, $r_{\exte}= 130$ mm. For
each Halbach cylinder the best configuration of the additional blocks have been marked
by a star. The black crosses in the figure are Halbach cylinder with the same
$r_{\inte}$ and $L$ as the Halbach cylinder with blocks, but with a bigger $r_{\exte}$
and no blocks. The black crosses are displaced by $\pm{} 6$ mm on the x-axis, to make
the comparison with the Halbach cylinders with additional blocks possible.}
\label{Fig.Magcool-parameter-extra-blocks}
\end{figure}

In Fig. \ref{Fig.Magcool-parameter-extra-blocks} $\Lambda_{\mathrm{cool}}$ is shown for
the different Halbach cylinders with additional blocks, i.e. the ones given in Table
\ref{Table.Halbach-cylinders-with-extra-blocks}. Here we can see that some
configurations of the additional blocks do increase $\Lambda_{\mathrm{cool}}$ by as much
as $\sim{}\hspace{-0.15cm}15\%$. Shown in the figure are also Halbach cylinders with no
additional blocks that have the same $r_{\inte}$ and $L$ as the Halbach cylinders with
additional blocks but where $r_{\exte}$ has been expanded by up to 30 mm. These are
shown such that it can be estimated if it is better to spend any additional magnetic
material on the additional blocks or on enlarging the external radius of the Halbach
cylinder. As one can see from the figure in three of the cases it is better to spend the
additional magnetic material on the additional blocks.

The design that is the most improved by the additional blocks is a short Halbach
cylinder with a large external radius, i.e. $L= 50$ mm, $r_{\exte}= 130$ mm. For
the longest Halbach cylinder with the smallest external radius, $L= 200$ mm, $r_{\exte}=
60$ mm, the additional blocks do not improve $\Lambda_{\mathrm{cool}}$. The
reason for this is that the short designs are also the ones that have the greatest loss
of flux through the ends of the cylinder bore, which is exactly what the additional
magnets reduce.

The optimal design of the additional blocks, i.e. the angular span, height and depth,
vary between the four Halbach cylinders presented in Fig.
\ref{Fig.Magcool-parameter-extra-blocks}, and the results can be read in Table
\ref{Table.Extra-blocks-optimal-magcool}. Here it is seen that the optimal design of the
additional blocks seems to be an angle span of around $60 ^{\circ}$, a height in the
range $45-60$ mm and a depth in the range $22.5-27.5$ mm, at least for the three systems
used in these simulations. The parameters were varied as given in Table
\ref{Table.Halbach-cylinders-with-extra-blocks}.

\begin{table}[!t]
\begin{center}
\caption{The optimal configuration of the additional blocks, i.e. the configurations
that maximize the magnetic cooling parameter. $(\Lambda_{\mathrm{cool}})_{\mathrm{org}}$
refers to the ``original'' Halbach cylinder without any additional blocks.}
\begin {tabular}{r|rll|c}
 Halbach                           & $\phi{}$  & $H$  & $D$ &  $\Lambda_{\mathrm{cool}}$\\
 dimensions                        &  &  &     &  $\overline{(\Lambda_{\mathrm{cool}})_{\mathrm{org}}}$\\
 $\textrm{[mm]}$                               &  [$^{\circ}$]    &  [mm]         &     [mm]     &   \\ \hline
 $r_{\exte}=130$, $L=50$         &  80  &      47.5 &     22.5 &  1.15\\\hline

 $r_{\exte}=130$, $L=92$          &  60  &      61.2 &     27.6 &  1.09\\ \hline

 $r_{\exte}=100$, $L=100$          &  60  &      50   &     25   &  1.03\\ \hline

 $r_{\exte}=60$, $L=200$          &  60  &      30   &     10   &  1.00\\ \hline

\end {tabular}
\label{Table.Extra-blocks-optimal-magcool}
\end{center}

\end{table}

It can thus be concluded that for a short Halbach cylinder with a large external radius
it is possible to optimize the magnetic cooling parameter by using additional magnets
placed at the ends of the cylinder. However, as can be seen by comparing Fig.
\ref{Fig.Magcool-parameter-Halbachs} and \ref{Fig.Magcool-parameter-extra-blocks} the
improvement gained by using the additional blocks is small compared to building a long
Halbach cylinder with a small $r_{\exte}$ in the first place. For example the design
improved the most by the additional blocks, $r_{\exte}=130$ mm and $L=50$ mm, has the
magnetic cooling parameter improved 1.15 times. This is not as impressive when one
considers that the magnetic cooling parameter for this Halbach cylinder has a value of
0.011 (the lower right corner in Fig. \ref{Fig.Magcool-parameter-Halbachs}), and even
multiplied by 1.15 this is still much lower than the longer Halbach cylinders. However,
in specific cases with a limited geometry due to the application the additional blocks
can still be used to improve the flux density.

\subsection{Homogeneity of the field}
In most Halbach cylinder applications it is not only the flux density that is important
but also the homogeneity of the field in the cylinder bore.

\begin{figure}[!t]
  \centering
   \includegraphics[width=\columnwidth]{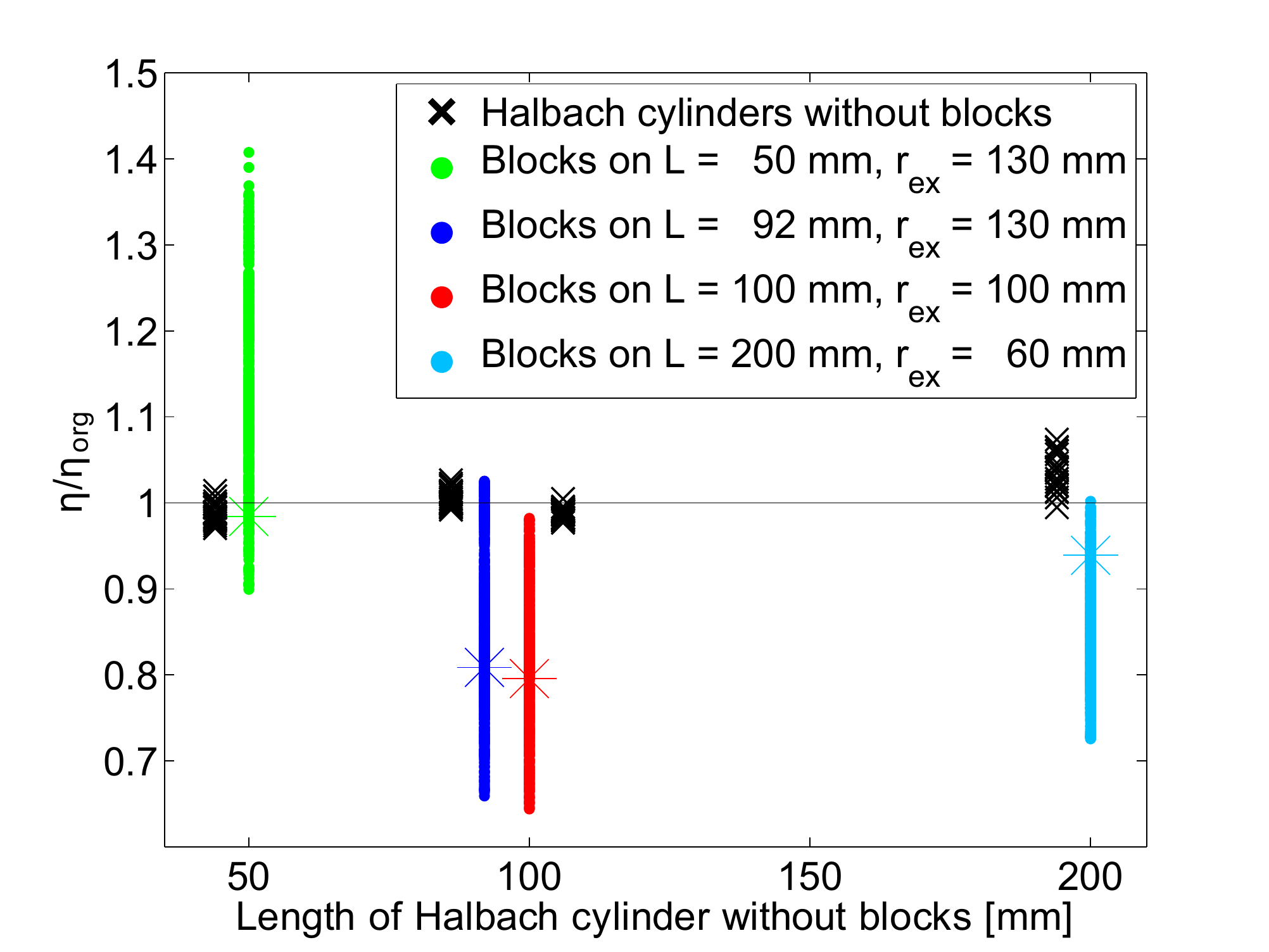}
      \caption{ The parameter $\eta/\eta_{\mathrm{org}}$ describing the homogeneity of
the field in the Halbach cylinder bore. The data point marked with a star represents the
best datapoint from Fig. \ref{Fig.Magcool-parameter-extra-blocks}. The black crosses are
Halbach cylinders with the same $r_{\inte}$ and $L$ as the Halbach with additional
blocks but with a larger $r_{\exte}$ and no blocks. The black crosses are displaced by
$\pm{} 6$ mm on the x-axis to make the comparison with the Halbach cylinders with
additional blocks possible.} \label{Fig.Homogenity-std-dev-extra-blocks}
\end{figure}

To characterizes the homogeneity of the flux density in the cylinder bore the quantity,
\begin{eqnarray}\label{Eq.Homogeneity-parameter}
\eta \equiv \frac{\avenotxt{2} - \avenoexp{}^2}{\avenotxt{2}},
\end{eqnarray}
where the angled brackets denoting volume average, is defined. In Fig. \ref{Fig.Homogenity-std-dev-extra-blocks} this parameter is shown
for the Halbach cylinders with additional blocks. Also shown in the figure are Halbach
cylinders with the same $r_{\inte}$ and $L$ as the Halbach with additional blocks but
with a larger $r_{\exte}$ and no blocks. It can clearly be seen that the no-block
designs with larger $r_{\exte}$ have a homogeneity parameter comparable to the ``original'' Halbach cylinder
without blocks, while a number of the designs with additional blocks clearly improve the
homogeneity of the field in the cylinder bore, i.e. lower
$\eta{}/\eta{}_{\mathrm{org}}$.

%-------------------------------------------------------------------------------------------------------------
%-------------------------------------------------------------------------------------------------------------
%-------------------------------------------------------------------------------------------------------------

\subsection{Direction of magnetization}
As previously mentioned the additional blocks all have a direction of magnetization
perpendicular to the end face of the Halbach cylinder. This might not be the optimal
configuration, so various directions of magnetization have been tested to find the
greatest enhancement of the flux density. The direction of magnetization was given by
$(0,B_r \textrm{cos}(\theta),B_r \textrm{sin}(\theta))$  where $\theta{}$ was varied in
steps of 1$^{\circ}$. The Halbach cylinder symmetry axis is oriented along the z-axis.

The result indicate that the mean flux density in the bore could only be improved by less
than 1\% by changing the direction of magnetization from the $90^{\circ{}}$ orientation
used in the preceding simulations.

\section{Discussion}
%------------------------------------ Revised article start -------------------------------------------
It is important to discuss the possible influence of the coercivity of the individual
magnets in the Halbach cylinder assembly. A serious problem in this context is that the
assumed linearity of the magnets is only valid when the magnetic field is above the
value of the intrinsic (polarization) coercivity, $H_{c}$. For typical 1.4 T NdFeB
magnets $\mu_{0}H_{c}$ is around 1.2 T at room temperature. Once the reverse component
of the magnetic field reaches this value the linearity of the magnets breaks down and a
small increase in the magnetic field will reverse the direction of magnetization of the
magnet. We have not modeled this nonlinearity but have assumed that the linear relation
is always valid. This is of course problematic when the magnetic field strength is too
high. The reason the nonlinearity is not modeled is that due to hysteresis a complete
history of the magnet material would be needed, including the physical building of the
Halbach array, and this is not possible to model.

\begin{figure}[!t]
  \centering
   \includegraphics[width=0.8\columnwidth]{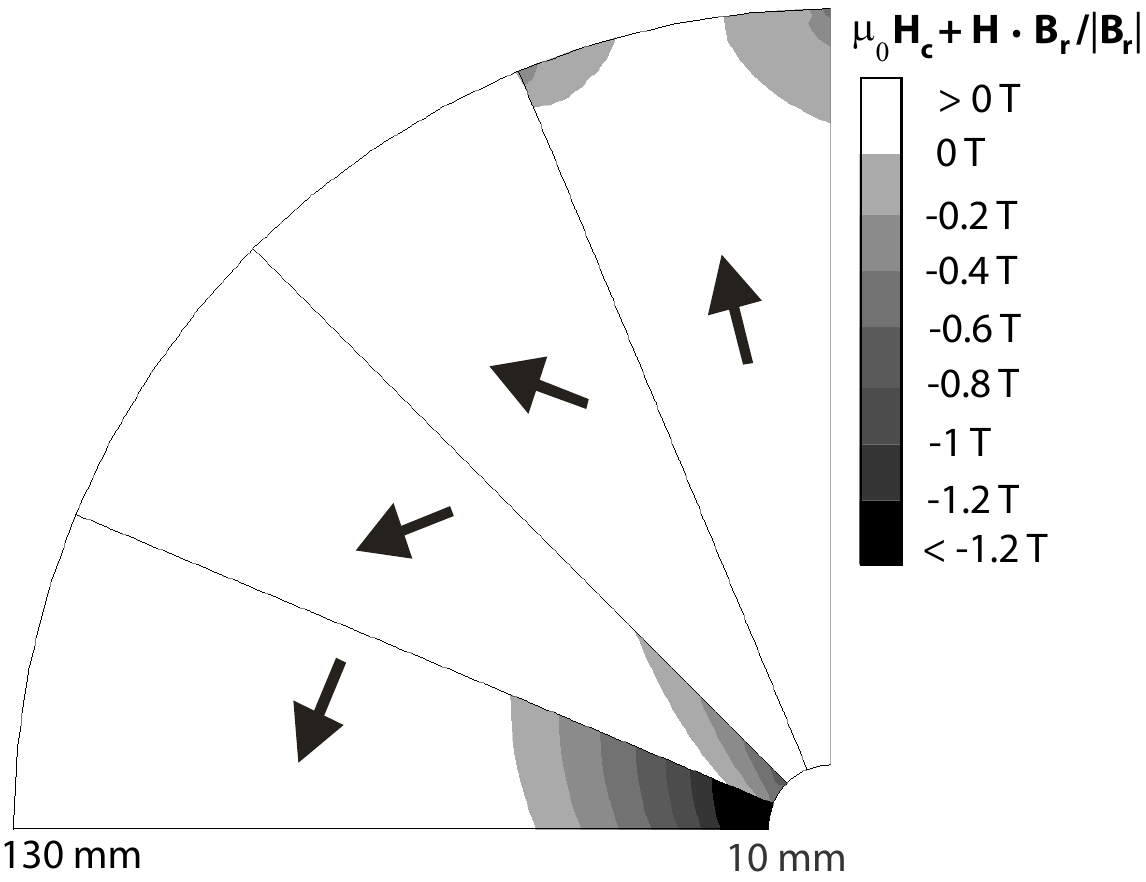}
      \caption{A quarter of 2D Halbach cylinder
       with $r_{\inte}=10$ mm and $r_{\exte}=130$ mm.
       Shown as a color map is the expression $\mathbf{\mu_0 H_c}+\mathbf{H}\cdot{}\mathbf{B_r}/|\mathbf{B_r}|$.
       When this expression is negative the magnetic field intensity is stronger than
       the intrinsic coercivity of the magnet (taken to be $\mu_{0}H_{c}=1.2$T) and the magnet will be reversed.
       The direction of the magnetization of the blocks are shown as black arrows.}
      \label{Fig.Coercivity}
\end{figure}

The part of the Halbach cylinder where this is a problem is around the inner equator\cite{Bloch_1998}.
In Fig. \ref{Fig.Coercivity} the
projection of the magnetic field intensity along the direction of the remanent magnetism
is shown for the largest Halbach considered in this paper.
Anisotropy is typically larger than coercivity for NdFeB magnets so any reverse field component is important \cite{Bloch_1998}.
It can be seen that the magnetic field is strong enough to reverse the direction of magnetization of typical
industry 1.4 T magnets at several locations.

The problem is present for Halbach cylinder with a flux density in the bore larger than 1.2 T.
To overcome this problem one can replace the magnets in the affected volume of the
Halbach cylinder with magnets with a higher intrinsic coercivity.
Such magnets are readily available at slightly lower remanences,
e.g. a typical industry NdFeB magnet with a remanence of
1.2 T has $\mu_{0}H_{c} = 3.2$ T, which is sufficiently strong to keep the relative
permeability constant and thus the material linear. Otherwise the best solution is to remove the magnets from this part of the Halbach
cylinder and replace them with a nonmagnetic material.

An additional remark on the conducted numerical simulations is that because the Halbach
cylinder consists of magnets with a relative permeability close to one, the
magnetostatic problem of calculating the flux density is linear in the remanence. This
means that the mean flux density both inside and outside the Halbach cylinder depends
linearly on the remanence of the magnets. In this paper we have used magnets with a
remanence of 1.4 T. If one would e.g. replace all these magnets in the Halbach cylinder
with magnets with a remanence of 1.2 T the mean flux density both inside and outside the
Halbach cylinder would decrease by a factor of $1.2/1.4 = 0.86$. This has be verified
numerically.

%------------------------------------ Revised article end ---------------------------------------------

There are still factors that have not been taken into account. We have for
example discussed the use of additional blocks while taking their shape for granted. It
is necessary to test if the circular design used for the additional blocks is the proper
design to use. One could just as well have used e.g. a square design of the additional
blocks. It is also important to investigate the effect of the additional blocks on a
much larger sample of Halbach designs, including designs with varying internal radii.

\section{Conclusion}
In this paper we found the optimal values of $r_{\exte}$ and $L$ for a Halbach cylinder
with a given mean flux density and $r_{\inte}$. These configurations have the smallest
volume of the magnet possible for a given mean flux density in the cylinder bore. Also,
we found that placing blocks of additional permanent magnets on the sides of the Halbach
cylinder can improve the flux density in the cylinder bore significantly. Finally, we
introduced a magnetic cooling efficiency parameter, $\Lambda_{\mathrm{cool}}$, and
showed that the additional blocks can improve this by as much as 15\% compared to
ordinary Halbach cylinders. However one must always take care that the polarization coercivity,
$H_{c}$, is always higher than the flux density in the Halbach cylinder gap.

\section*{Acknowledgements}
The authors would like to acknowledge the support of the Programme Commission on Energy
and Environment (EnMi) (Contract No. 2104-06-0032) which is part of the Danish Council
for Strategic Research.

%$ $Id: Halbach.tex,v 1.39 2008/04/25 11:49:14 rbjk Exp $ $}

\end{document}